\documentclass[aps,prl,10pt,showpacs,twocolumn]{revtex4-1}

\usepackage[utf8]{inputenc}
\usepackage{amsfonts}
\usepackage{amsmath}
\usepackage{amssymb}
\usepackage[utf8]{inputenc}

\usepackage{graphicx}
\usepackage{subfigure}

\begin{document}

\title{Plasmonic Graded-Chains as Deep-Subwavelength Light Concentrators.}

\author{Natalia Esteves-L\'opez${}^{1}$, Horacio M. Pastawski${}^{2}$, Ra\'{u}l A. Bustos-Mar\'{u}n${}^{1,2}$}
\pacs{73.20.Mf, 73.21.-b, 78.67.-n, 42.79.Ek}

\affiliation{${}^{1}$Departamento de Fisicoqu\'{\i}mica, Facultad Ciencias Qu\'{\i}micas, UNC, Ciudad Universitaria, 5000 C\'{o}rdoba, Argentina, ${}^{2}$IFEG, Facultad de Matem\'{a}tica Astronom\'{i}a y F\'{i}sica, UNC, Ciudad Universitaria, 5000 C\'{o}rdoba, Argentina}

\email{rbustos@famaf.unc.edu.ar}

\begin{abstract}
We have studied the plasmonic properties of aperiodic arrays of identical nanoparticles (NPs) formed by two opposite and equal graded-chains (a chain where interactions change gradually). We found that these arrays concentrate the external electromagnetic fields even in the long wavelength limit.
The phenomenon was understood by identifying the system with an effective cavity where plasmonics excitations are trapped between effective band edges, resulting from the change of passband with NP's position.
Dependence of excitation concentration on several system's parameter was also assessed. This includes, different gradings as well as NP's couplings, damping, and resonant frequencies. In the spirit of the scaling laws in condensed matter physics, we developed a theory that allows us to rationalize all these system's parameters into universal curves. The theory is quite general and can also be used on many other situations (different arrays for example). Additionally, we also provided an analytical solution, in the tight-binding limit, for the plasmonic response of homogeneous linear chains of NPs illuminated by a plane wave.
Our results can find applications on sensing, near field imaging, plasmon-enhanced photodetectors, as well as to increase solar cell efficiency.
\end{abstract}

\maketitle

\section{INTRODUCTION.}

Concentration or focusing of electromagnetic fields (EMFs) has been a goal of great interest almost since the inventions of lenses. This is due to its potential applications, which in modern times span from imaging and sensing to lasing. Traditionally mirrors, lenses, or combinations of them, such as resonant cavities \cite{cavity,Fainstein}, have been used for this purpose. But the diffractive nature of electromagnetic waves has always imposed a hard limit to these devices.
There have been several proposals to overcame this, including superlenses made of negative refractive index metamaterials \cite{PendryNRI,ShalaevNRI}, superoscillations \cite{superoscillation}, and time reversal focusing \cite{Fink,stockmanTR}. However, difficulties associated with them, such as complex designs involving active materials or the need for an almost absolute control of the incident EMF, have undermined their development into concrete applications.

Probably the most successful strategy for subwavelength, or even deep-subwavelenght, concentration of light is the exploitation of evanescent waves. All though predicted a long time ago \cite{Mie,synge}, we had to wait until the end of the 20th century, when we acquired enough control of matter at nanoscale, to see those predictions turn into experimental results \cite{SNOM,First,CoronadoSchatz,BrongersmaRev,Reinhard}.
Despite its success, there are however certain drawbacks in using evanescent waves for light concentration, as for example the volume-versus-confinement problem.

It is known that certain nanostructure presents under illumination regions of high EMFs or hotspots. Now, if one want to increase even further the EMFs inside a hotspot one can sharpen the tips or the region of the nanoparticle (NP) where concentration takes places. However, this also reduces the volume of the hotspots \cite{Perassi}, an effect that can not be neglected if the target molecule to be sensed is a macromolecule for example, or the goal is to increase the efficiency of solar cells \cite{AtwaterPolmanRev}.

Another possibility, involves the use of arrays of NPs or nanoscopic structures where radiation concentration occurs inside the system as consequence of the interaction among its different components \cite{BrongersmaRev}.
Here, two different situations should be distinguished.
The first one occurs when nanostructures have one or more dimensions approaching the excitation wavelength and thus it is necessary to consider retardation effects. The concepts behind many of these retardation-based plasmonic light concentrators are based on scaled radio frequency antenna designs \cite{BrongersmaRev,HernandezExp,GarciaAntena,Maier,NovotnyAntena}.
There are of course other alternatives, such as the use of aperiodic metallic waveguide arrays \cite{Aperiodic}.
The second situation occurs when the size of the nanostructure is significantly smaller than the wavelength of the incident light. Here, the entire structure experiences a uniform electric field at any instant and then the quasistatic approximation is valid.
Examples of this are given by the use of fractal aggregates \cite{fractal}or self-similar chains \cite{nanolens} of subwavelength particles usually called nanolenses.
It is worth noting that in the mentioned and in many other examples, the lack of periodicity of the arrays seems to be key in EMF's focusing.

Aperiodic arrays of NPs, sometimes called graded arrays if the parameters describing the array vary in a graded way, have shown interesting and uniques features. Such as the appearance of new types of excitations called gradons \cite{GradonsYu}, or a continuous frequency-depend localization of excitations in different part of the system \cite{GradedMalyshev}.

\begin{figure}
\begin{center}
\includegraphics[width=3.3in]{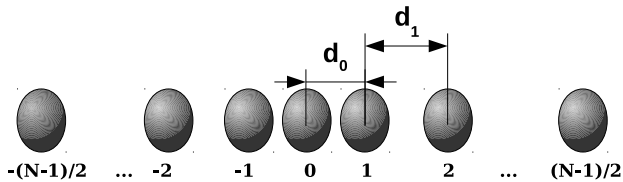}
\end{center}
\caption{Scheme of the system studied, a mirrored graded-chain of NPs. It consists of two opposite and equal graded-chain of identical metallic NPs.}
\label{system}
\end{figure}
In this work we study a kind of aperiodic array of NPs, that we called mirrored graded-chain of NPs. They are formed by two equal but opposite graded-chains of NPS, see Fig. \ref{system}. Array's geometry allows it to concentrate external EMFs, but with the important property that excitation concentration does not come at the expense of reducing hotspot's volume, as in the case of nanolenses.
Through the work we focus on the physics behind the phenomenon and the effect of different system's parameters on excitation concentration.
In the spirit of the scaling laws in condensed matter physics, we develop the theory that allows one to build universal curves for the properties studied. The theory, valid also for other arrays of NPs, requires only arrays made of equal NPs that can be treated in the quasistatic limit.

The work is organized as follows:
In section I we briefly introduce the method used for the calculations. In section II we develop various aspects of the  behavior of linear chains of equidistant NPs that will result useful in the next sections.
In section III we present the main results of our work. This section is subdivided into two main subsections ``Nearest-neighbor approximation'' and ``Dependence on system's parameters``. In the first one we use a simple version of our system to extract the physics behind excitation concentration. While in the second one, we systematically study  different aspects that could affect that. In this last subsection we develop a simple theory that allows us to build universal curves for the frequency of the maximum excitation concentration and the strength of it. There is also a short subsection at the end ''Comparison with simple graded-chains'' where we briefly compare excitation concentration between mirrored and simple graded-chains.
Finally, in section IV we summarize the main conclusions.

\section{Coupled dipole approximation.}
In this work we study linear arrays of NPs where the interactions among them are modeled using the coupled dipole approximation \cite{BrongersmaCDA,Citrin,BurinSchatz,Markel,BCP-01,BCP-02,BDCP}. In this model, each $n^{th}$-NP is described by a dipole moment $P_n$ induced by the electric field produced by the other dipoles $m$ at $n$, $E_{n,m}$, and the external electric field at $n$, $E_n^{ext}$. Assuming a generic ellipsoidal shape for the NPs, this gives for the induced dipole moment \cite{CoronadoSchatz,BCP-02,BDCP}:
\begin{equation}
\vec{P}_{n}=
\frac{\epsilon _{0}V(\epsilon -\epsilon _{\mathrm{med}})}{\left[
\epsilon _{\mathrm{med}}+L(\epsilon -\epsilon _{\mathrm{med}})\right] }
\vec{E}_{n}^{\mathrm{total}}
\label{Pn},
\end{equation}
where $\vec{E}_{n}^{\mathrm{total}}=\vec{E}_{n}^{(ext)}+ \sum_{m\neq n}^{N} \vec{E}_{n,m}$,
$\epsilon_{0}$ is the free space permittivity, $V$ is the volume, $\epsilon$ is the dielectric constant of NP's material,  $\epsilon_{\mathrm{med}}$ is the dielectric constant of the host medium, and $L$ is a geometric factor that depends on NP's shape and the direction of $E$ relative to the array \cite{CoronadoSchatz}.

For a linear array of NPs, transversal and longitudinal excitations, $T$ and $L$ respectively, do not mix, which allows us to write $E_{n,m}$ as \cite{BCP-02,BDCP}:
\begin{equation}
E^{T/L}_{n,m}=- \frac{\gamma^{_{T/L}}_{n,m} P^{T/L}_{m}}{4\pi \epsilon _{0}\epsilon_{\mathrm{med}}d_{n,m}^{3}}, \label{Enm},
\end{equation}
where $d_{n,m}$ is the distance between NPs, and the complex constant $\gamma$ depends on the wavenumber $k$ of the excitation and the orientation of the NP's array relative to the direction of $P$.
\begin{align}
& \gamma^{_{T}}_{n,m}=[1-ikd_{n,m}-(kd_{n,m})^{ 2 }]e^{ ikd_{n,m} }  \notag \\
& \gamma^{_{L}}_{n,m}=-2[1-ikd_{n,m}]e^{ ikd_{n,m} }. \label{gTL}
\end{align}
Note that in the quasistatic approximation, $kd \approx 0$, $\gamma^{_{T}}$  and $\gamma^{_{L}}$ are $1$ and $-2$ respectively.

The dipolar moments $P_n$, and the external electric fields $E_n^{ext}$ can be arranged as vectors $P$ and $E$ resulting in:
\begin{equation}
\mathbf{P} = \left( \mathbb{I}\omega ^{2}-\mathbb{M}\right)^{-1} \mathbb{R} \mathbf{E}=
\mathbf{\chi} \mathbf{E},
\label{matrix}
\end{equation}
where $\chi$ is the response function, $\mathbf{M}$ is the dynamical matrix and $\mathbf{R}$ is the diagonal matrix that rescales the external applied field according to the shape, volume and material of the NP.

Assuming a particular model for $\epsilon$ in Eq. \ref{Pn} one can obtain the expressions for the different elements of $M$ and $R$. For ellipsoidal NPs and using a Drude-Sommerfeld's like model for $\epsilon$,
$\epsilon = \epsilon_{\infty}- \omega_{_\mathrm{P}}^{2}/(\omega ^{2}+i\omega \eta )$,
one obtains \cite{BCP-02,BDCP}
\begin{equation}
M_{n,n}=\widetilde{\omega }_{_\mathrm{SP}(n)}^{2}=\omega_{_\mathrm{SP}(n)}^{2} - i \Gamma_n(\omega), \label{OmegaSP}
\end{equation}
\begin{equation}
M_{n,m}=- \omega_{\mathrm{x}(n,m)}^{2} = - E^{T/L}_{n,m} R_{n,n},
\label{OmegaX}
\end{equation}
and
\begin{equation}
R_{n,n}=-\epsilon _{0}V_{n} \omega_{_\mathrm{P}(n)}^{2} f.\label{Rnn}
\end{equation}
where $\omega_{_\mathrm{P}(n)}$ is the plasmon frequency of the NP with index ``$n$'', $\omega_{_\mathrm{SP}(n)}^{}$ is its resonant frequency, $\Gamma_n(\omega)$ is its decay rate, and $\omega_{\mathrm{x}(n,m)}^2$ is the coupling between NPs $n$ and $m$.
The factor $f$ can in principle depend on $\omega$, for 
$\epsilon_{\infty} \neq \epsilon _{\mathrm{med}}$, which typically slightly shifts the resonances of the system, see Ref. \cite{BCP-02,BDCP}. However, as we are mainly interested in extracting general trends of our system, we will take $f$ as constant for simplicity. For the same reason, we will consider only the case of equal NPs with the same resonant frequency for transversal and longitudinal excitations, and a linear form for $\Gamma(\omega)$, i.e. $\Gamma(\omega)=\eta \omega$, which neglects radiation damping.

According to the above expressions, the role of $\omega_{_\mathrm{SP}}$ is in general to set the scale of the problem, which is reflected as a shift in the spectra or in the dispersion relations.
For this reason and in order to gain generality, from now on we will only use rescaled variables for the
excitation frequencies, coupling constants, damping terms, $R_{ii}$, and the self energy $\Pi$ (see next section) of the linear arrays. That is, those variables should be understood as $\omega'=\omega/\omega^{}_{_\mathrm{SP}}$, $\omega'^{}_{\mathrm{x}}=\omega^{}_{\mathrm{x}}/\omega^{}_{_\mathrm{SP}}$, $\eta'=\eta/\omega^{}_{_\mathrm{SP}}$, $R'_{ii}=R_{ii}/\omega^{2}_{_\mathrm{SP}}$, and $\Pi'=\Pi/\omega^{2}_{_\mathrm{SP}}$ respectively.

\section{Homogeneous linear chain of nanoparticles.} \label{SecHomogeneous}
\begin{figure}
\includegraphics[width=2.5in, trim=0.5in 0.0in 0.2in 0.1in, clip=true]{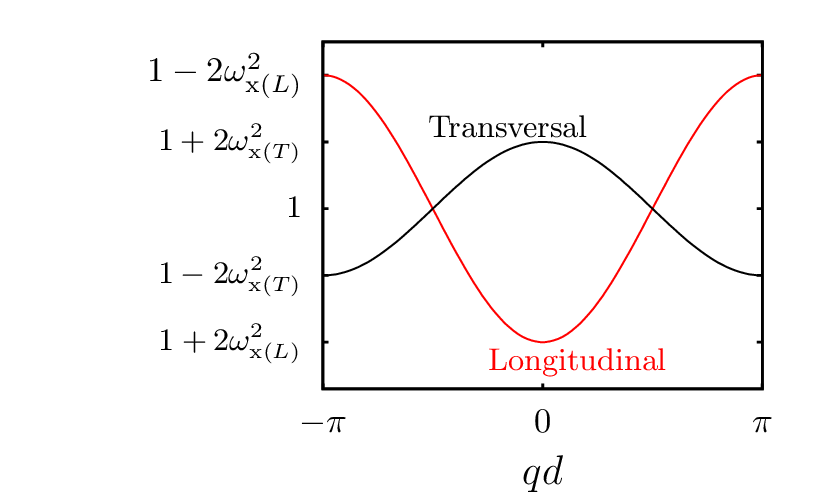}
\caption{Typical dispersion relation for longitudinal and transversal excitations of NP's homogeneous-linear-chains. Calculations were done assuming only nearest-neighbor interactions, $\eta=0$,  and a quasistatic approximation.}
\label{FIGdispersion}
\end{figure}

In this section we discuss different aspects of the behavior of homogeneous linear chains of NPs, a system where distances among neighboring NPs are equal. The discussions here presented will result useful in the next sections.
In order to obtain simple analytical expressions that will allow us to understand our problem in simple terms, we will treat our systems in the quasistatic approximation and neglect interactions beyond first neighbors.

For an homogeneous linear chain, the dipolar moment of two NPs can only differ in a phase factor, thus
$P_n=P_m e^{-(m-n) qd}$ where $d$ is the distance between neighboring NPs. Considering only nearest-neighbor interactions and negligible damping terms, one can find the value of the wave vector $q$ for a given frequency of excitation, i.e. the dispersion relation of our system \cite{BrongersmaCDA,BCP-01},
\begin{equation}
\omega^{2}(q)=1 + 2\omega_{\mathrm{x}}^{2} \cos(qd). \label{dispersion}
\end{equation}
Fig. \ref{FIGdispersion} shows an example of this, where it can be seen that changing the type of plasmonic excitation not only shrinks the passband, the interval of frequencies for which Eq. \ref{dispersion} has solution, but it also inverts the curve's shape due to a change in the sign of $\omega_{\mathrm{x}}^{2}$. The maximum and minimum values of the passband are known as band edges and for this particularly simple example are given by:
\begin{equation}
\omega_{\pm}^{2}(q)= 1 \pm 2\omega_{\mathrm{x}}^{2}. \label{bandedge}
\end{equation}

Now, let us analyze the excitation of a linear array of NPs illuminated by a plane wave. The electric field of the external source at NP ``n'' is
\begin{equation}
\vec{E}_n^{(ext)}=E_0 \hat E e^{i n k_z d}  \label{Eext}
\end{equation}
where $E_0$ is the modulus of $\vec{E}$, $\hat E$ its direction, $n=(...,-1,0,1,2...)$ the NP's index, and $k_z d = k d \cos(\theta)$ with $\theta$ being the angle between the NP's chain and the incident light.
In a homogeneous linear chain of NPs with only nearest-neighbor interaction one can calculate analytically all the elements of the response function $\chi$, defined in Eq. \ref{matrix} \cite{BCP-01,Cattena-02}
\begin{equation}
\chi_{n,m} = \chi_{n,n} \prod^{m-1}_{l=n} \left ( \frac{\Pi_l}{\omega^2_{\mathrm{x}}} \right )  \label{ChiIJ}
\end{equation}
where all the self energies $\Pi_l$'s are equal in the limit of infinite NPs ($N\rightarrow\infty$)
\begin{align}
\Pi (\omega )& =\tfrac{1}{2}\left[ \omega ^{2}- 1 + i \eta \omega \right] -  \label{Self} \\
& \mathrm{sgn}(\omega ^{2}-1)\tfrac{1}{2}\sqrt{\left[
\omega ^{2}- 1 + i \eta \omega \right] ^{2}-4\omega_{\mathrm{x}}^{4}},  \notag
\end{align}
and 
\begin{equation}
\chi _{n,n}=\frac{R_{n,n}}{\left[ \omega ^{2}-1 + i \eta \omega \right] -2 \Pi (\omega )}.  \label{chinn}
\end{equation}
Therefore, when the whole chain is illuminated, the dipolar moment $P_n$ of a given NP can be calculated from:
\begin{align}
P_n^{(T/L)} &=\sum_m \chi_{n,m} E^{ext(T/L)}_m \notag \\ 
& =  \left ( \chi_{n,n} E_0^{(T/L)} \right ) \left \{ 1 + \sum_{m=1}^{\infty} \left( \frac{\Pi}{\omega_{\mathrm{x}}^2} e^{i k_z d} \right )^m \right. \notag \\  
& \left. + \sum_{m=1}^{\infty} \left( \frac{\Pi}{\omega_{\mathrm{x}}^2} e^{- i k_z d} \right )^m \right \}  \label{PiChi},
\end{align}
which yields
\begin{equation}
\frac{P_n^{(T/L)}}{E_0^{(T/L)}}= \chi_{n,n}
\left\{
\frac{2 \left [ 1- \left(\frac{\Pi}{\omega_{\mathrm{x}}^2}\right) \right ] }
{1-2\left(\frac{\Pi}{\omega_{\mathrm{x}}^2}\right) +\left(\frac{\Pi}{\omega_{\mathrm{x}}^2}\right)^2} -1
\right \}.
\end{equation}
where $E_0^{(T/L)}$ is the transversal or one of the longitudinal components of $E_0 \hat E$.

\begin{figure}
\begin{center}
\includegraphics[width=3.3in]{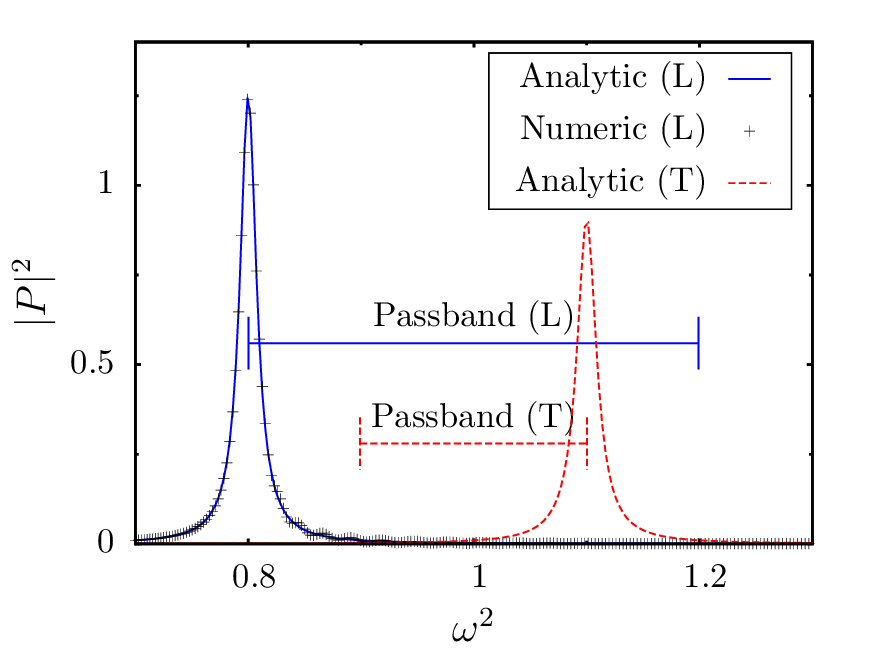}
\end{center}
\caption{Square dipolar moment, $|P|^2$, of a NP in a homogeneous-linear-chain as function of the square frequency of the external field, $\omega^2$. The wave vector of the external field is perpendicular to the chain with $E$ longitudinal (L) or transversal (T) to it. The nearest-neighbor approximation as well as the quasistatic approximation are considered, $\eta=0.01$ and $\omega_{\mathrm{x}}^2=-0.1$ and $0.05$ for L and T excitations respectively.
All curves are normalized to the maximum value of $|P|^2$ of an isolated NP with $\eta = 0.01$. Numerical results correspond to $|P|^2$ of the central NP of a finite array of 81 NPs.}
\label{FIGTB-Homogeneous}
\end{figure}

Fig. \ref{FIGTB-Homogeneous} shows an example of a linear homogeneous array of NPs illuminated perpendicular to the chain. As can be seen, the analytical solution
overlaps the numerical results. The maximum of each spectrum occurs at one of the band edges, the lower one for longitudinal excitations, negative $\omega_{\mathrm{x}}^2$, and the higher one for transversal excitations, positive $\omega_{\mathrm{x}}^2$. This behavior can be understood in terms of Eq. \ref{PiChi} in the limit of small $\eta$. For example in the case of longitudinal excitations the fraction $\left(\frac{\Pi}{\omega _{_\mathrm{x}}^2}\right)$ is exactly ``1'' at the lower band edge and exactly ``-1'' at the higher band edge.
That means that two consecutive terms of Eq. \ref{PiChi} contribute constructively in the former case and destructively in the latter one (the other way around for transversal excitations). 
In more physiccal terms, that means that the absorption of light by two consecutive nanoparticles interfere contructively or destructively if the excitation frequency is one or the other band edge.
Furthermore, the density of states has peaks only at the band edges, so only there it is expected a considerable absorption of 
the external field. In Fig. \ref{FIGTB-Homogeneous}, we can also notice that the height of the peaks for T and L excitations are approximately the same as that of
an isolated NP, note that the normalization factor is precisely the maximum value of $|P|^2$ of an isolated NP. The small differences in the peaks' high are due to the frequency dependence of the damping terms. This is also reasonable as all NPs are being excited and absorb in the same way and the resulting excitation
must be distributed without a preferential direction through the array.

Finally, it is useful to briefly discuss the effect of several deviations from the ideal model considered above. As can be found in several references
\cite{BrongersmaCDA,Citrin,BurinSchatz,Markel,BCP-01,BCP-02,BDCP}, the effect of taking into account all orders of interparticle couplings is essentially a widening of the passband plus
a shifting of its center. Retardation effects on the other hand, have a stronger influence on the properties of the chain, driving the appearance of new resonances and valleys product of constructive and destructive interferences in the interparticle couplings. Those effects are stronger at higher
frequencies and when the size of the problem (NP's radii or interparticle distances) is not negligible compared with the wavelength of the excitation.
However, in this work we will focus on systems with the appropriate scale such as to safely neglect these retardation effects. At the end of the last section we
will retake this issue. The presence of defects in the chain, such as NPs of different radii or inhomogeneous NP's separations, induces the appearance of
localized states, which correspond to collective excitations of a reduced number a NPs whose resonance generally lies outside the passband \cite{BCP-01}.

\section{Mirrored graded-chain of nanoparticles.}

Now that we have discussed some relevant aspects of linear arrays of equidistant NPs, we can focus on understanding the effect of changing from homogeneous interactions to slightly graded couplings. As shown in Fig. \ref{system}, the system studied consists of an inhomogeneous linear array of metallic NPs where the distance between neighboring NPs incresases from the center to the ends of the array while the size and shape of NPs are kept constant.
Two different models for distance's increasing are studied
\begin{equation}
\begin{array}{lcl}
d_n=d_0 \alpha^{|n|}  & \ \ &\mathrm{model \ 1 \ (Exponential)} \\
d_n=d_0 (1+ |n| \beta)  & \ \ &   \mathrm{model \ 2 \ (Linear)}
\end{array} \label{model-alpha}
\end{equation}
where $\alpha$ and $\beta$ are the incremental factors, $n=0$ corresponds to the central NP, and we have simplified the notation for $d$ by replacing $d_{n,n+1}$
(or $d_{-n,-(n+1)}$) by $d_n$ (or $d_{-n}$).

We will see that this type of systems focus the external electromagnetic field on the central NPs.
In what follows we will first analize our system in the nearest-neighbor approximation. Then, we will disscus the effect of taking into account different aspects that approach our simple model to a more realistic situation, but always
trying to find general trends in its behavior that can result useful for designing
potential applications.

\subsection{Nearest-neighbor approximation.}

\begin{figure}
\begin{center}
\includegraphics[width=2.9in]{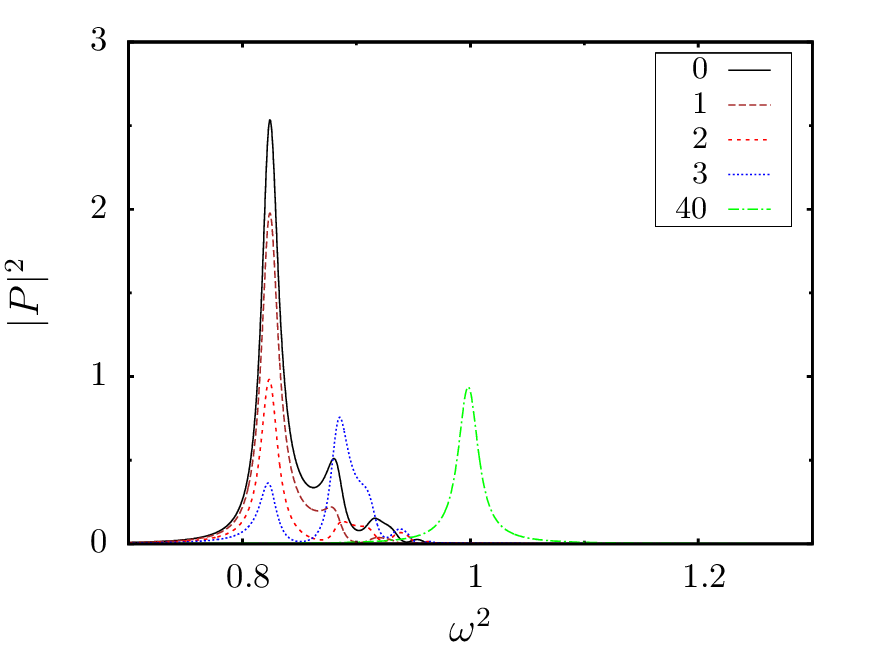}
\end{center}
\caption{$|P|^2$ vs $\omega^2$ for different NPs in a mirrored graded-chain of NPs (see inset). Nps' distances are changed according to model 1 (Eq. \ref{model-alpha}) with $\alpha=1.035$. The system is illuminated perpendicularly to the array with $E$ longitudinal to the chain. Nearest-neighbor approximation as well as the quasistatic approximation are considered. We used $\omega_{\mathrm{x}}^2=-0.1$, $\eta=0.01$, and the total number of NPs is 81.
All curves are normalized to the maximum value of $|P|^2$ of an isolated NP with $\eta=0.01$.}
\label{FIGSpectrum-TB-Graded}
\end{figure}

Fig. \ref{FIGSpectrum-TB-Graded} shows an example of the ``spectrum'' (where spectrum stands for $|P|^2$ vs $\omega^2$)
of the studied system. As we can see, now the maximum of the ``spectrum`` of the central NP is slightly shifted from the band edge position toward the center of the passband.
New resonances appear, especially for NPs far from the center, but more interestingly, there is a considerable increase of the peak's height of the central NPs together with a decrease of it for NPs away from the center. Indeed, not only the height of the spectrum is increased but also its total area.

\begin{figure}
\begin{center}
\includegraphics[width=3.3in, trim=0.1in 0.0in 0.1in 0.1in, clip=true]{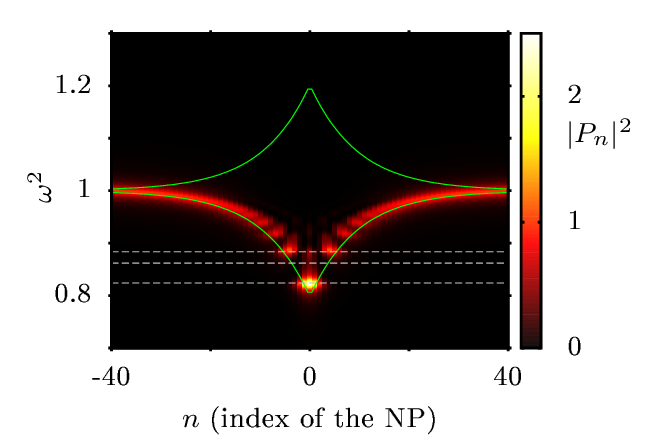}
\end{center}
\caption{(Color online) - 3D plot of $|P|^2$ vs $\omega^2$ and NP's position in the array. Conditions are the same as in Fig. \ref{FIGSpectrum-TB-Graded}.
Continuous green lines show the band edges of the local equivalent homogeneous chains (see text and Eq. \ref{bandedge}). Grey dashed lines mark the three lowest eigenvalues of $\mathbb{M}$ for longitudinal excitations.}
\label{FIG3D-TB-Graded}
\end{figure}

Fig. \ref{FIG3D-TB-Graded} shows a 3D color map of $|P|^2$ vs $\omega^2$ and NP's position. The plot presents a typical seagull's contour with well resolved resonances around the middle that slowly turns into a continuum toward the edges.
The general shape of the plot can be understood by following the approach of treating the system as ``locally'' homogeneous \cite{GradonsYu,Aperiodic}.
That means that even though the properties of the array change continuously with NP's position, these changes are smooth enough to treat the system locally as an homogeneous linear array.
Thus, we can define a position dependent band edge at every NP's position. 
The continuous green lines of Fig. \ref{FIG3D-TB-Graded} mark exactly that. Note, the almost perfect matching between the green line and the maximum of the excitation for NPs away from the center.

Xiao et.al. \cite{GradonsYu} studied infinite linear arrays of NPs and harmonic oscillators with graded interactions in a given direction, which would be equivalent to take only one half of our system (see Fig. \ref{system}). The key difference between their and our case is that here an excitation at a given frequency can not travel indefinitely in a given direction, even with $\eta=0$. For example, take an excitation that start at the crossing between the grey dashed line at the highest frequency and the descending green line that correspond to the lowest passband edge (the lower left one). This excitation can travel through the chain only to the right (propagation to the left will put the excitation outside the passband). This excitation can continue traveling but only until it reaches the other crossing with the band edge, the one at the right. Beyond that point the excitation would be again outside the passband.
This implies that excitations are necessarily trapped between the crossing with the band edges and thus the system acts as an effective optical cavity.

As it is well known, optical cavities can increase enormously electromagnetic fields inside them due to coherent accumulation of excitations. This explains not only the concentration of excitations
in our system but also the discrete nature of resonances around the center of the system. Obviously, due to finite damping,
excitations that has to travel distances greater than a characteristic decay length to interfere with themselves will not present this ``optical cavity'' effect. This explains why the system behaves as homogeneous linear chains for NPs far apart from the center.

\begin{figure}
\begin{center}
\includegraphics[width=2.0in, trim=0.0in 0.0in 0.0in 0.0in, clip=true]{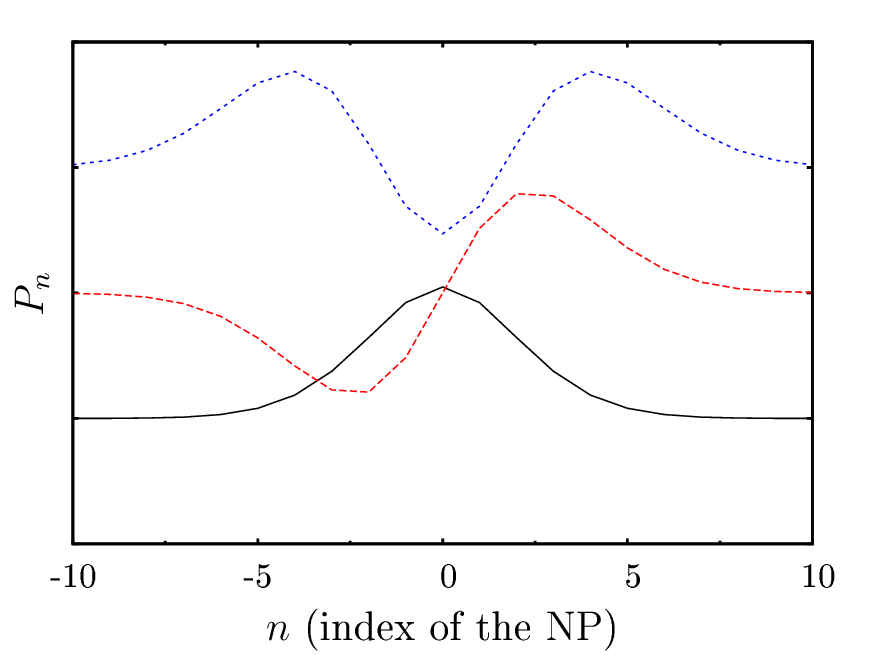}
\end{center}
\caption{The three first normal modes of a mirrored graded-chain of NPs. Conditions are the same as in Fig. \ref{FIGSpectrum-TB-Graded}.}
\label{FIGModes}
\end{figure}

Calculation of cavity's resonant frequencies is not that simple in our system as the effective cavity changes its length with frequency and its group velocity with position. However one can always calculate numerically the eigenvalues of $\mathbb{M}$. Grey lines in Fig. \ref{FIG3D-TB-Graded} show the real part of the three lowest eigenvalues of $\mathbb{M}$. Those are the values at which the resonant condition occurs. Naturally not all of them are bright modes as the excitation field, a plane wave perpendicular to the array, can not excite anti-symmetric modes for example.
The eigenmodes corresponding to those eigenvalues are shown in Fig. \ref{FIGModes}. Note that the lowest eigenmode has, as expected, its maximum at the center of the array which explains why the highest concentration of radiation occurs around the central NP.
For transversal excitation, modes are similar but $P_n$ changes its sign between neighboring sites. The spectra are also similar as well as the 3D plots but, as expected, the peaks for the different NPs appear close to their highest band edges (with $|\omega^2_{\mathrm{x}(L)}/2|=|\omega^2_{\mathrm{x}(T)}|$). Besides that, the only difference is that peak's heights are lower due to a higher damping ($\Gamma=\eta \omega$).

\subsection{Dependence on system's parameters.} \label{Deviations}

In this section, we will consider different factors that deviate the system from the simple model discussed previously.
In all the figures shown, we have taken $|P_0|_{\mathrm{max}}^2$, the maximum value of $|P|^2$ of the central
NP, as an indicator of excitation concentration. Another possibility could have been, for example, to take the area under the curve of the spectra of $|P_0|^2$ vs $\omega^2$. However, as the results proved to be qualitatively the same, we will not discuss the last one.

\begin{figure}
     \begin{center}
        \subfigure{
            \includegraphics[width=3.3 in, trim=0.0in 0.0in 0.1in 0.0in, clip=true]{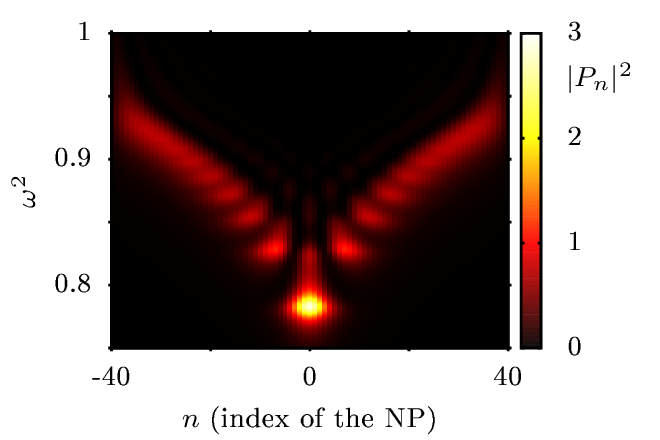}
        }        
                \\
        \subfigure{
            \includegraphics[width=1.6 in, trim=0.0in 0.0in 0.3in 0.1in, clip=true]{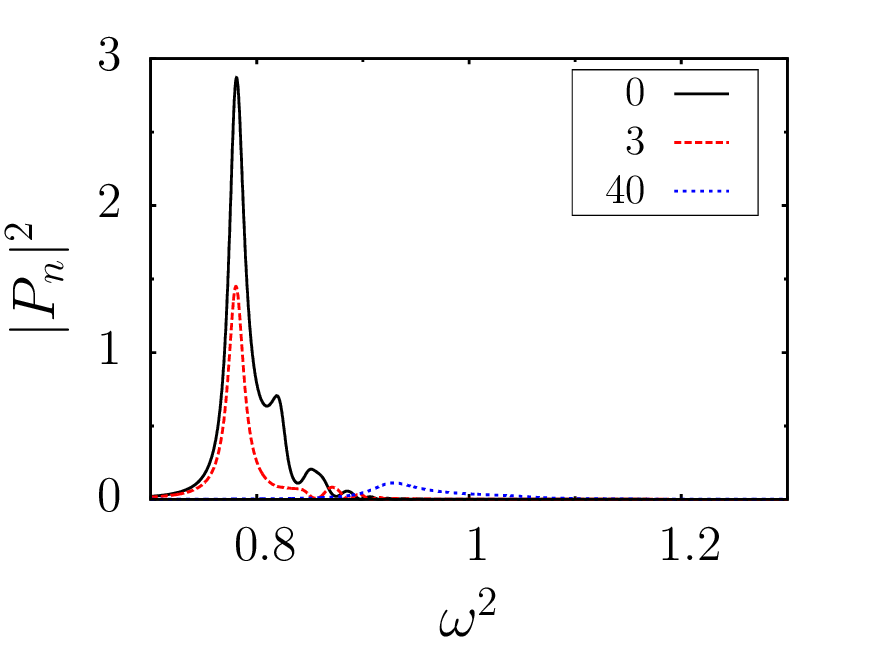}
        },\subfigure{
            \includegraphics[width=1.6 in, trim=0.0in 0.0in 0.3in 0.1in, clip=true]{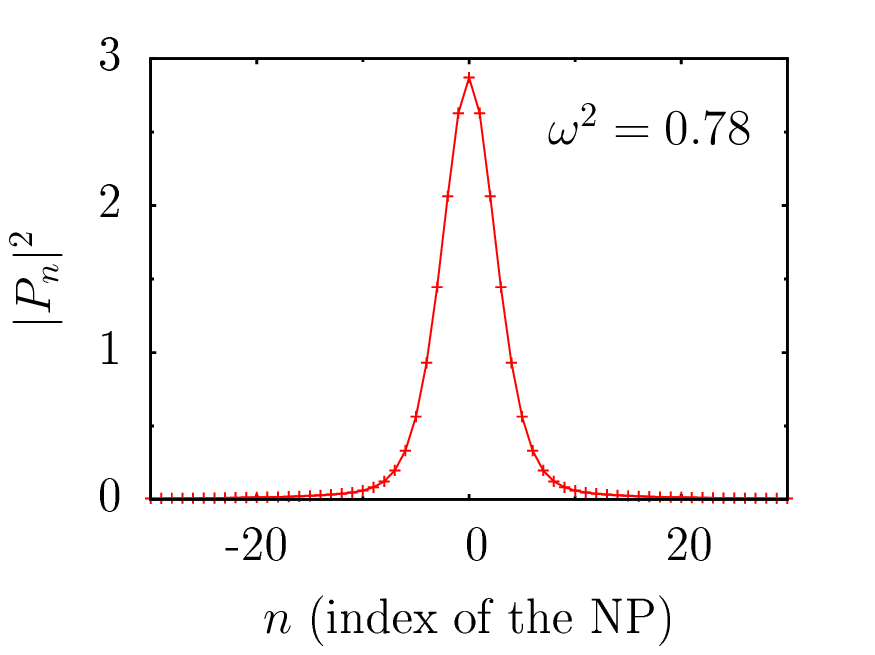}
        }
    \end{center}
    \caption{\textbf{Upper figure} - The same as Fig. \ref{FIG3D-TB-Graded} but $\alpha=1.01$ and all interactions are considered. \textbf{Lower left figure} - Vertical cuts of the upper figure at $n=0$, $3$, and $40$ (same  as Fig. \ref{FIGSpectrum-TB-Graded} but under the present conditions). \textbf{Lower right figure} -  Horizontal cut of the upper figure at $\omega^2=0.78$ (response of the system at constant frequency of excitation).}
   \label{FIG3D-Graded}
\end{figure}

Fig. \ref{FIG3D-Graded} shows an example of the response of a graded-chain of NPs under the same condition as in Fig. \ref{FIG3D-TB-Graded} but considering all interactions among NPs, not only first neighbors.
Qualitatively the behavior of the system is essentially the same.
The differences are that the concentration of radiation is slightly increased and the peak's positions are shifted towards lower frequencies, both are better appreciated in the lower left sub-figure. The shifting is expected as the band edges of NP's waveguides are red-shifted when considering all interactions among NPs and longitudinal excitations\cite{BrongersmaCDA,Citrin,BurinSchatz,Markel,BCP-01,BCP-02,BDCP}.
Apart from that, the only difference is that NPs of the end of the chain (NP's index -40 and 40 in the plot) do not show yet the response of isolated NPs as in Fig. \ref{FIG3D-TB-Graded}.
This is also reasonable as the effective interaction between these NPs and the rest of chain is larger when we consider interaction beyond first neighbors. Thus, a larger array would be required to observe that.

The lower right sub-figure of Fig. \ref{FIG3D-Graded} shows the square dipolar moment of different NPs of the chain for a fixed frequency of excitation. Note, how quickly excitation concentration decays when moving away from the central NPs. This effect is interesting because the system, which is a 1-D planar structure, behaves similarly to the tips of a SNOM though \cite{SNOM}. A phenomenon that, in principle, can be used for near field spectroscopies.

\begin{figure}
\begin{center}
\includegraphics[width=2.9in, trim=0.0in 0.0in 0.0in 0.0in, clip=true]{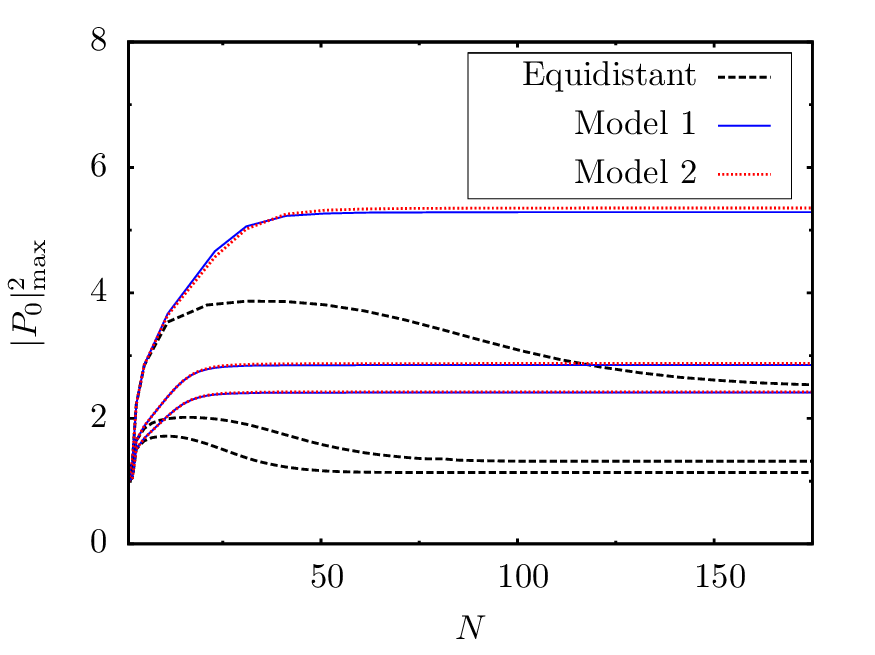}
\end{center}
\caption{Maximum value of $|P|^2$ of the central NP of arrays with different length ($N$).
Three values of $\omega_{\mathrm{x}}^2$ are used: -0.25, -0.1 and -0.05 with $\eta=0.01$.
blue continuous lines, mirrored graded-chains where distances are incremented exponentially with $\alpha$ equal to 1.005, 1.02, and 1.02. Red dotted lines, the same but distances are incremented linearly with $\beta=1-\alpha$. Black dashed lines, homogeneous linear chains of NPs. Lowest curves correspond to smallest $\omega_{\mathrm{x}}^2$.}
\label{FIGP2-N}
\end{figure}
\begin{figure}
\begin{center}
\includegraphics[width=2.9in, trim=0.0in 0.0in 0.0in 0.0in, clip=true]{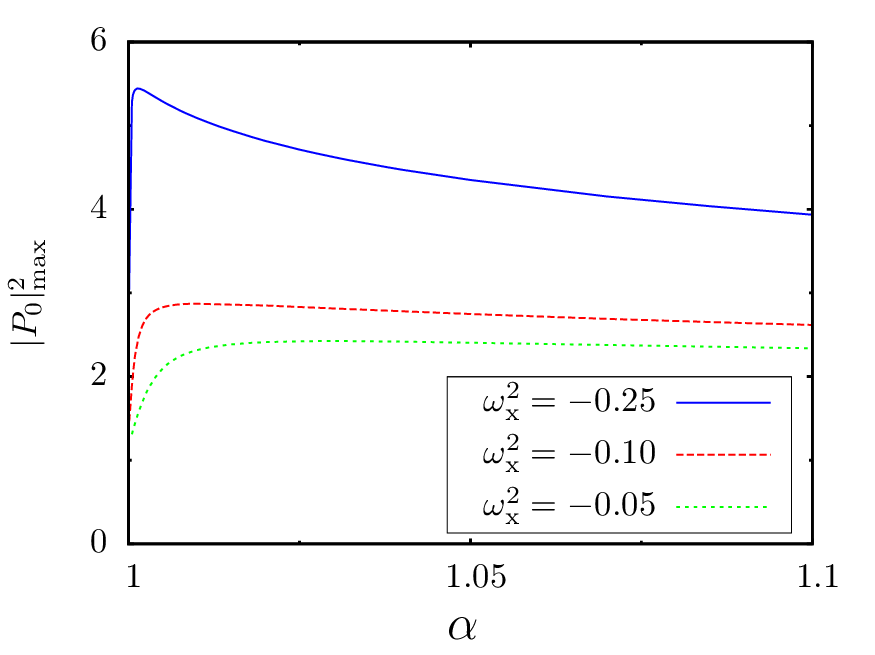}
\end{center}
\caption{Maximum value of $|P|^2$ of the central NP of a mirrored graded-chain as function of $\alpha$ (distances change exponentially with ``n''). Calculations are converged with respect to $N$ and $\eta=0.01$.}
\label{FIGP2-Alpha}
\end{figure}

Fig. \ref{FIGP2-N} is an example of the convergence of $|P_0|_{\mathrm{max}}^2$, the maximum value of $|P|^2$ of the central NP, with the total number of NPs in the array. Note that a graded-chain of only 20 or 40 NPs, depending on $\omega_{\mathrm{x}}^2$, shows a better concentration of external excitation than any finite or infinite 
chain of equidistant NPs. Interestingly, both models for graded-chains, see Eq. \ref{model-alpha}, yield almost the same result for the values of $\alpha$ and $\beta$ used. Those values correspond roughly to the optimal ones (see Fig. \ref{FIGP2-Alpha}). The reason for that is simple. Even though a graded-chain of NPs is always better than a chain of equidistant NPs, the optimal one corresponds surprisingly to an array of almost equidistant NPs, i.e. $\alpha$ and $\beta$ close to 1 and 0 respectively. Expanding $d_n/d_0$ around $\alpha=1$ for the exponential model, gives
\begin{equation}
\alpha^{n} \approx 1 + n (\alpha-1) + O(\alpha^2).
\end{equation}
Thus,
\begin{equation}
\left.  d_n(\alpha) \right |_{\mathrm{model 1}} \approx \left.  d_n(\beta = \alpha-1) \right |_{\mathrm{model 2}}
\end{equation}
For that reason, hereafter we will exclusively discuss the results for $d_n$ given by one of them, the exponential model.

Let us now discuss the effect of $\alpha$ on excitation concentration. It is clear from Fig. \ref{FIGP2-N} that there should be an optimal value of $\alpha$ between $1$ and $\infty$, as $\alpha=1$ correspond to equidistant NPs and $\alpha=\infty$ to an isolated NP. Fig. \ref{FIGP2-Alpha} shows precisely that. However, the optimal value of $\alpha$ is very close to 1 as discussed. Indeed, the bigger the coupling ($\omega_{\mathrm{x}}^2$), the closer to 1 results this optimal value.

\begin{figure}
        \includegraphics[width=2.9 in, trim=0.0in 0.0in 0.1in 0.0in, clip=true]{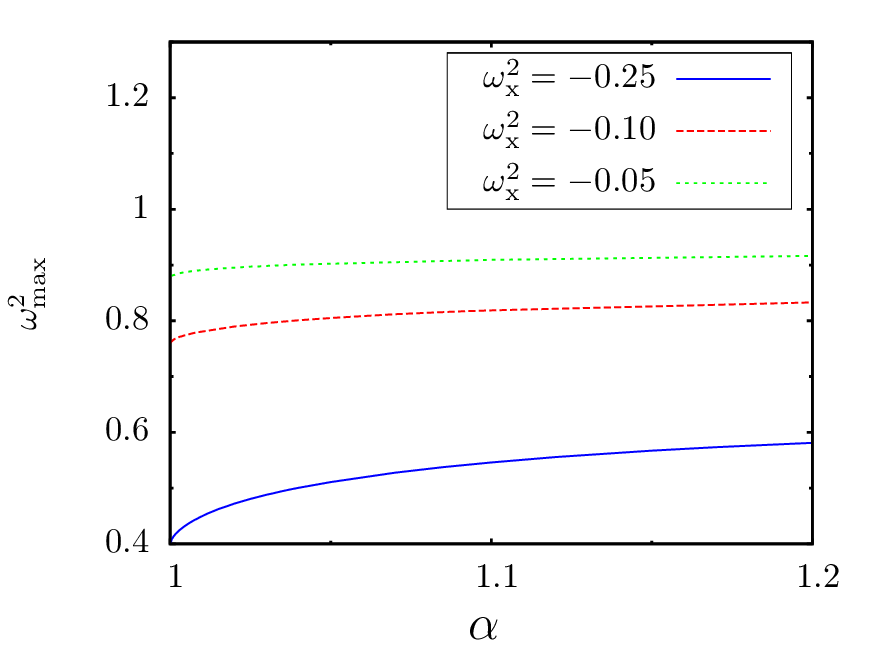}         
        \caption{Position of the spectrum's peak of $|P|^2$ vs $\omega^2$, $\omega^2_{\mathrm{max}}$, for the central NP of a mirrored graded-chain as function of $\alpha$. Calculations are converged with respect to $N$ and $\eta=0.01$.}
       \label{FIGw-Alpha}
\end{figure}

\begin{figure}
       \includegraphics[width=2.9 in, trim=0.0in 0.0in 0.1in 0.1in, clip=true]{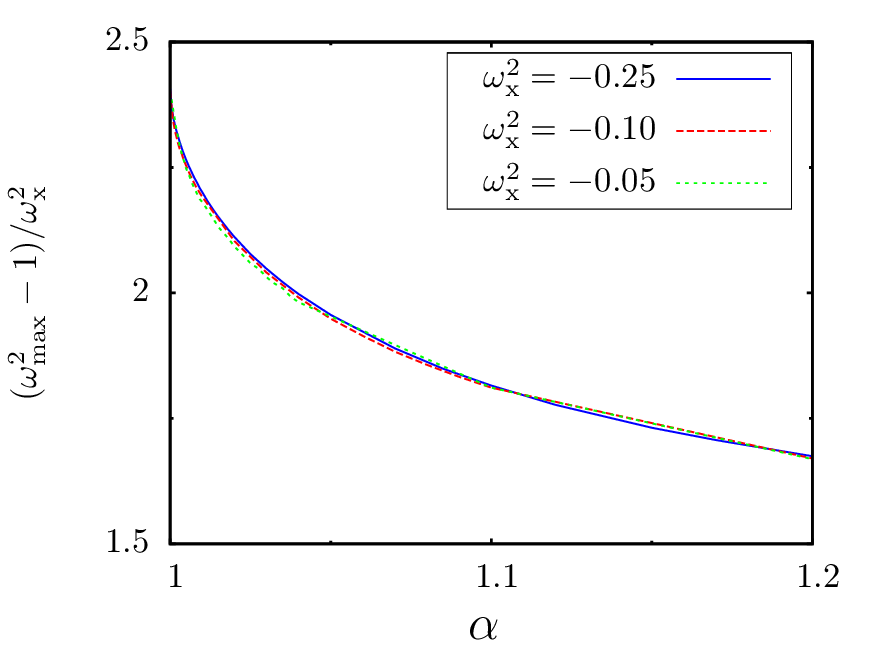}
        \caption{Same as Fig. \ref{FIGw-Alpha} but using a rescaled square resonant frequency of the central NP.}
       \label{FIGw-Alpha-Universal}
\end{figure}

The position of the peak of $|P_0(\omega)|^2$, $\omega^2_{\mathrm{max}}$, also changes with $\alpha$. At $\alpha=1$, it corresponds to the position of the lower band edge of the equivalent homogeneous NP's chain. At $\alpha=\infty$ it is $1$, which corresponds to the value of an isolated NP. Fig. \ref{FIGw-Alpha} shows this behavior. As can be seen in the figure different couplings give different curves. However, all curves can be rescaled to a single universal one by means of a simple linear transformation, see Fig. \ref{FIGw-Alpha-Universal}. This behavior can be understood by analyzing how the eigenvalues of the $\mathbb{M}$ matrix, Eq. \ref{matrix}, change with $\alpha$. First, we have to rewrite the matrix $\mathbb{M}$ as:
\begin{equation}
\mathbb{M} = \omega_{\mathrm{x}}^2 \mathbb{M}' + (1 - i \Gamma(\omega))\mathbb{I}
\end{equation}
where the matrix $\mathbb{M}'$, defined through its elements as $\mathbb{M}'_{n,n}=0$ and $\mathbb{M}'_{n,m}= \left (d_0/d_{n,m}\right )^3$, depends only on the geometry of the array. For example for $d_n$ given by the exponential model, $\mathbb{M}'_{n,m}=1/\left( \sum_{l=n}^{m-1} \alpha^{|l|}\right )^3$. Clearly a matrix $\mathbb{U}$ that diagonalize $\mathbb{M}$ also diagonalize $\mathbb{M}'$ which implies:
\begin{equation}
\mathbb{D} = \omega_{\mathrm{x}}^2 \mathbb{D}' + ( 1 - i \eta \omega) \mathbb{I}
\end{equation}
where $\mathbb{U}^{-1} \mathbb{M} \mathbb{U}=\mathbb{D}$ and $\mathbb{U}^{-1} \mathbb{M}' \mathbb{U}=\mathbb{D}'$, being $\mathbb{D}$ and $\mathbb{D}'$ the diagonal matrices of the eigenvalues of $\mathbb{M}$ and $\mathbb{M}'$ respectively.
Therefore, the eigenvalues of $\mathbb{M}$ which give the position of system's response maxima are given by the eigenvalues of a matrix that only depends on the geometry of the array ($\alpha$ in our case) and $\omega_{\mathrm{x}}^2$ through
\begin{equation}
\omega_{\mathrm{max}}^2(\omega_{\mathrm{x}}^2,\alpha)= \omega_{\mathrm{x}}^2 \mathbb{D}'_{0/N}(\alpha) + 1 \label{Universalw}
\end{equation}
where $\mathbb{D}'_{0/N}$ is the lowest eigenvalue of $\mathbb{M}'$ for longitudinal excitations or the highest one for transversal excitations, and it depends only on $\alpha$ in our case.

\begin{figure}
\begin{center}
\includegraphics[width=2.9in, trim=0.0in 0.0in 0.0in 0.0in, clip=true]{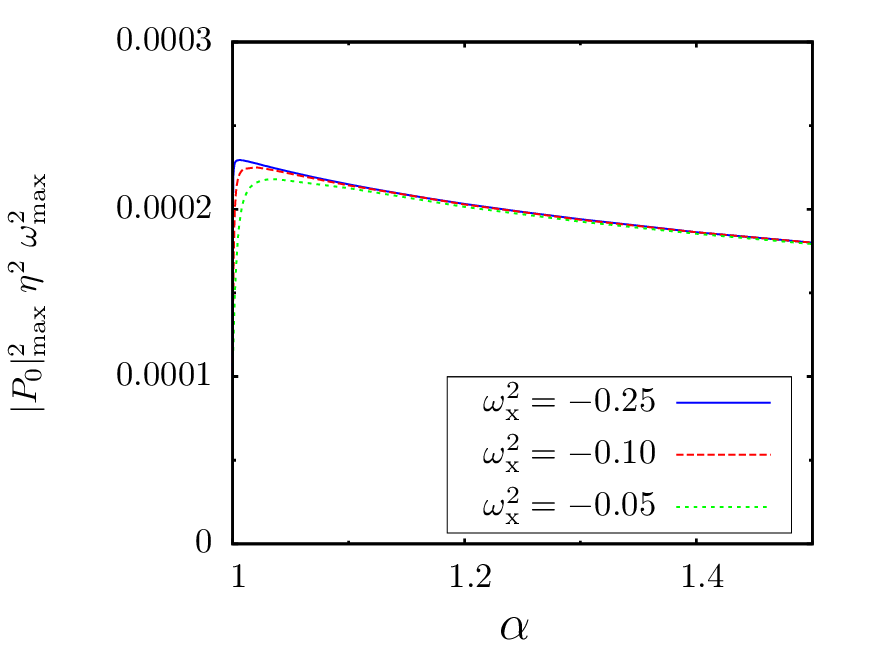}
\end{center}
\caption{Same as Fig. \ref{FIGP2-Alpha} but using a rescaled maximum square dipolar moment of the central NP.}
\label{FIGP2-Alpha-Universal}
\end{figure}

Similarly, one can find the universal behavior of $|P_0|^2$ by noticing that the eigenvectors of $\mathbb{M}$ also depend only on the geometry of the array. Using Eq. \ref{matrix}, we can write
\begin{equation}
P_n= \sum_{l,m} U_{n,l} \frac{1}{\omega^2-D_{l,l}} U_{l,m}^{-1} R_{m,m} E^{(ext)}_m
\end{equation}
Now, as all NPs are the same $R_{m,m}=R_0$ and the system can be treated in the quasistatic approximation $E^{(ext)}_m=E_0$, then
\begin{equation}
\frac{P_n}{R_0 E_0}= \sum_l \frac{U_{n,l}(\alpha)  \sum_m U_{l,m}^{-1}(\alpha)  }{\omega^2-
\left ( \omega_{\mathrm{x}}^2 D'_{l,l}(\alpha) + 1 - i \eta \omega \right )
}. \label{Pgeneral}
\end{equation}
By using this equation one can in principle calculate any property of interest of our system by knowing its geometric contributions, $U$ and ($D'$), and the rescale parameters $R_0$, $\omega_{\mathrm{x}}^2$, and $\eta$ which do depend on the material, size and shape of individual NPs. Under certain conditions the expression can be simplified even further.
For example, one can assume that the main contribution to the above summation at $\omega^2=\omega^2_{\mathrm{max}}$ is given by $l=0/N$. Then the following relation holds
\begin{equation}
\frac{ |P_0(\omega_{\mathrm{x}}^2,\eta,\alpha)|^2_{\mathrm{max}} }{|R_0|^2 |E_0|^2} = \frac{ \left| U_{0,0/N} \sum_m^{} U_{0/N,m}^{-1} \right |^2 } {\eta^2 \omega^2_{\mathrm{max}}} \label{UniversalP}
\end{equation}
where the numerator on the right hand side depends only on the geometry of the array ($\alpha$ in our case). Fig. \ref{FIGP2-Alpha-Universal} shows the validity of this equation for different values of $\omega_{\mathrm{x}}^2$ and $\alpha$. We can see that the convergence is excellent especially for large values of $\alpha$ or high couplings.

By interpreting the system as an effective cavity, we can easily understand the deviations of Eq. \ref{UniversalP}.
For ($\alpha$)s close to $1$, the curves that represent the band edges as function of NP's positions, see for example the continuous green line of Fig. \ref{FIG3D-TB-Graded}, becomes flater which will make the eigenvalues of $\mathbb{M}$ to collapse.
The curvature of these lines depend on both $\omega_{\mathrm{x}}^2$ and $\alpha$. The smaller the $\alpha$ and/or $\omega_{\mathrm{x}}^2$ the less pronounced the curvatures and the less separated the eigenvalues of $\mathbb{M}$. Therefore, it is reasonable to see deviation of Eq. \ref{UniversalP} for $\alpha$ close to $1$ and small $\omega_{\mathrm{x}}^2$.

\begin{figure}
\begin{center}
\includegraphics[width=2.9in, trim=0.0in 0.0in 0.0in 0.0in, clip=true]{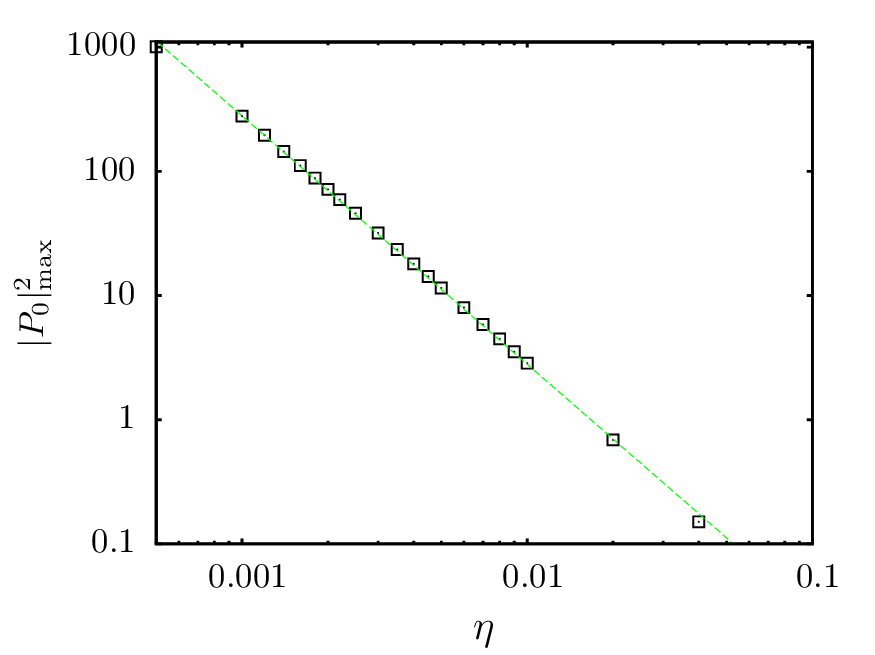}
\end{center}
\caption{Maximum square dipolar moment of the central NP of a mirrored graded-chain as function of $\eta$. $\alpha=1.02$ and $\omega^2_{\mathrm{x}}=-0.1$. Green line is a fit to a $\eta^{-2}$ function.}
\label{FIGP2-Eta}
\end{figure}

Fig. \ref{FIGP2-Eta} shows an example of the variation of $|P_0|^2_{\mathrm{max}}$ with $\eta$ which gives an almost perfect $1/\eta^2$ dependence as expected. At large ($\eta$)s, compared with the separation among eigenfrequencies, the approximation in Eq. \ref{UniversalP} obviously breaks down which is the origin of the small deviations at large ($\eta$)s.
 
It is interesting to highlight that Eqs. \ref{UniversalP} and \ref{Universalw} (or some equivalent) are valid beyond the particular model treated here of mirrored graded-chains. In particular Eq. \ref{Universalw} (or the equivalent one for other $P_i$s and its corresponding $D_{l,l}$s) is valid for any system of equal NPs in the quasistatic approximation, while Eq. \ref{UniversalP} also requires values of $\eta$ smaller than the eigenfrequencies' separation, i.e. well defined resonances.
Retardation effects affects $\mathbb{D}$ and the numerator of Eq. \ref{UniversalP}. Thus, in a sense, they modify the geometry or the effective geometry of the array. Regretfully this turns the clean linear relation between $\mathbb{D}'_{0/N}$ and $\omega^2_{\mathrm{max}}$ into a complex non-linear problem as now 
$\mathbb{D}'\equiv\mathbb{D}'(\alpha,k_0d_0,\omega)$, see Eq. \ref{k0d0}.
Anyway, provided the necessary conditions are met, Eqs. \ref{Universalw} and \ref{UniversalP} (or even Eq. \ref{Pgeneral} in the case of not well resolved resonances)  can be used to find universalities in the behavior of any system of interest even if we can not solve it analytically.


\begin{figure}
     \begin{center}
        \subfigure{
            \label{fig:first}
            \includegraphics[width=2.9 in, trim=0.0in 0.0in 0.1in 0.0in, clip=true]{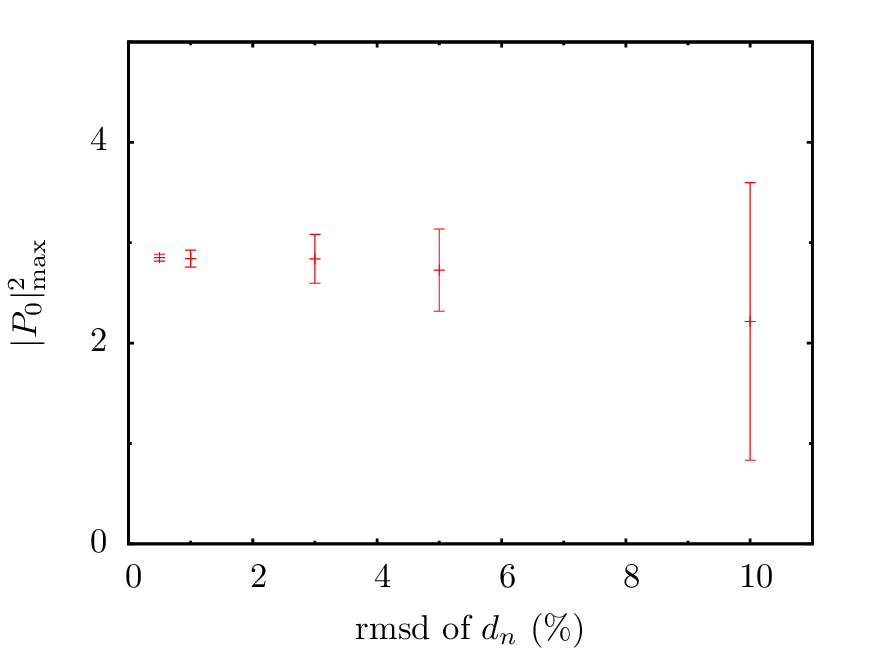}
        }        
        \\
        \subfigure{
            \label{fig:fourth}
            \includegraphics[width=2.9 in, trim=0.0in 0.0in 0.1in 0.1in, clip=true]{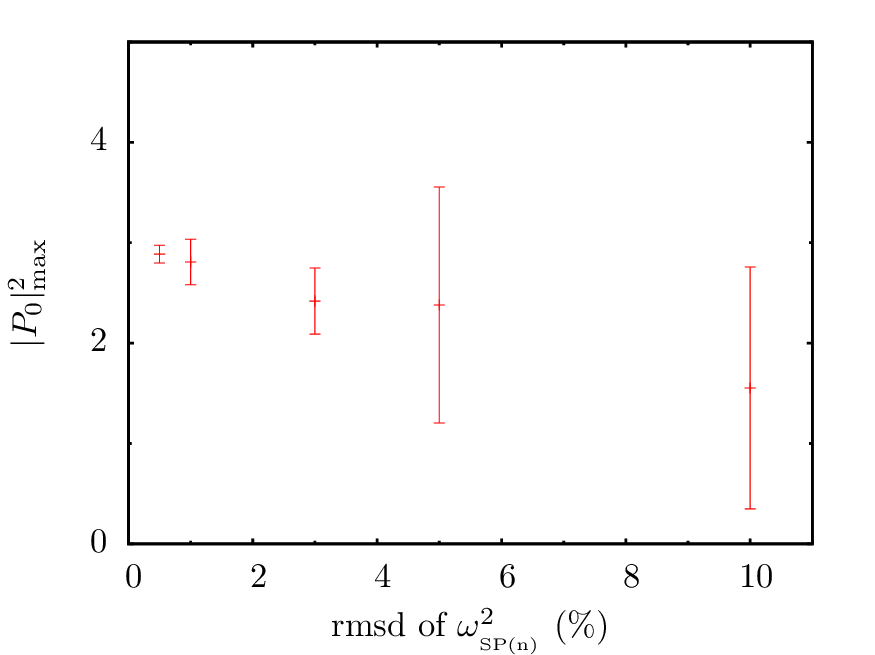}
        }
    \end{center}
    \caption{\textbf{Upper figure} -Average value and root-mean-square deviation (rmsd) of $|P_0|^2_{\mathrm{max}}$ for mirrored graded-chains with random defects in $d_n$. We used a normal probability distribution to sample the random defects in $d_n$. The X-axis shows the rmsd of $d_n$ and the height of the vertical bars the rmsd of the values of $|P_0|^2$ obtained. $\alpha=1.02$, $\omega^2_{\mathrm{x}}=-0.1$,and $\eta=0.01$. -\textbf{ Lower figure} - The same but considering random defects on the resonant frequency of individual NPs.}
   \label{FIGNoise}
\end{figure}

Returning to the particular system studied here, it is important to address the question of how sensitive is our proposed structure to defects in the array. Fig. \ref{FIGNoise} shows how $|P_0|^2$ changes when imperfections are introduced into the array.
In the upper figure we changed randomly the distances between NPs ($d_n^{\mathrm{new}}=d_n^{\mathrm{old}}+\Delta d $) by using a normal distribution to sample $\Delta d /d_n^{\mathrm{old}}$.
In the lower figure we did the same but changing NP's resonance frequency.
The width of the normal distribution used to sample the distances (or resonant frequencies) is what is indicated in the x axis as ``rmsd of $d_n$'' (``rmsd of $\omega_{\mathrm{SP(n)}}^2$ ''). We run 10 calculations for every condition and calculate the average value of $|P_0|^2$ and
its standard deviation, which is indicated as an error bar in the plots. We can see that the system exhibits a good fault tolerance. For up to 3\% or 5\% of noise the energy concentration remains almost the same.
Being, the fault tolerance to defects in the distances ($d_n$) better than that to the resonant frequencies ($\omega_{\mathrm{SP(n)}}^2$).
This tolerance to defects approaches our proposal to its implementation. Furthermore, it is important to highlight that when energy concentration is based on retardation effects, systems result, in general, very sensitive to defects in its components.

\begin{figure}
\begin{center}
\includegraphics[width=2.9in, trim=0.0in 0.0in 0.0in 0.0in, clip=true]{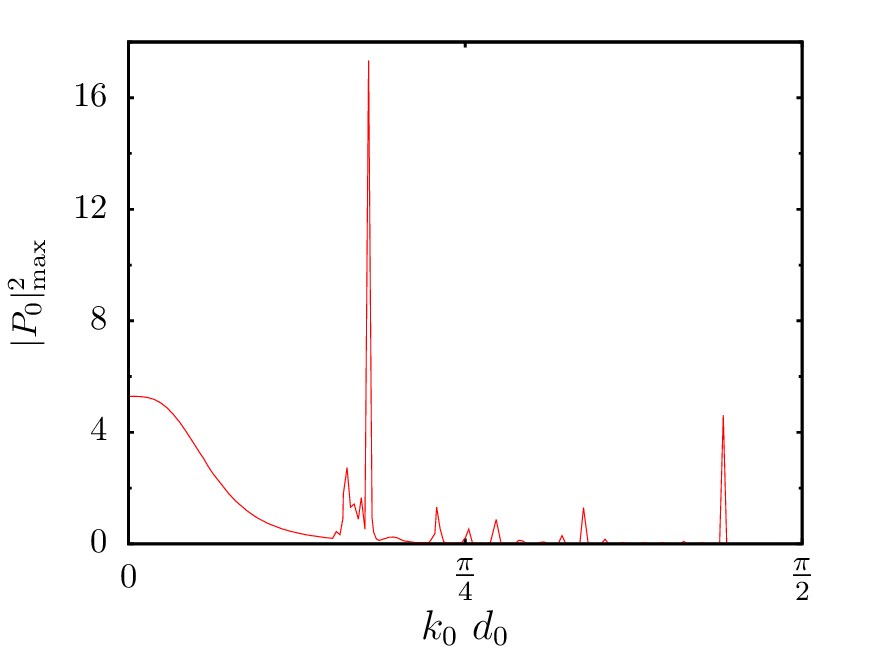}
\end{center}
\caption{Maximum square dipolar moment as function of retardation effects (see Eq. \ref{k0d0}) for the central NP of a mirrored graded-chain with $\alpha=1.005$, $\omega^2_{\mathrm{x}}=-0.25$,$\eta=0.01$.}
\label{FIGRetardations}
\end{figure}

Finally, in order to study retardation effects but without resorting to particular examples, with a given value of $d$, $\omega_{\mathrm{SP}}$, etc..., we define $k_0$ as.
\begin{equation}
k_0=\omega_{\mathrm{SP}}/c 
\end{equation}
where $c$ is the velocity of light, and rewrite $kd$ in Eq. \ref{gTL} as
\begin{equation}
kd=\omega (d_{n,m}/d_0) (k_0 d_0) \label{k0d0}
\end{equation}
where $\omega$ is as usual the renormalized excitation frequency (the true $\omega$ divided by $\omega_{_\mathrm{SP}}$) and $k_0 d_0$ is the parameter that will account for retardation effects.
Fig. \ref{FIGRetardations} shows an example of the consequences of increasing retardation effects. 
If $k_0d_0$ is still small, we see a plateau and then a smooth decreasing of energy concentration on the central NP. When retardation effects increase even further there are, as expected, very narrow resonances at some precise values of $k_0d_0$ that produce a high concentration of the external excitations. However, the resonances are so sensitive to the specific value of $k_0$ and $d_0$ that any defect in the distances or a shift in the frequency of excitation would destroy it. Furthermore, at a fixed frequency of excitation, increasing $k_0d_0$ is equivalent to enlarge the whole system which is contrary to our goal, to develop nanostructures that focus external fields in the smallest possible region.

\subsection{Comparison with simple graded-chains.} \label{simpleGraded}

\begin{figure}
\includegraphics[width=2.7in, trim=0.0in 0.0in 0.1in 0.1in, clip=true]{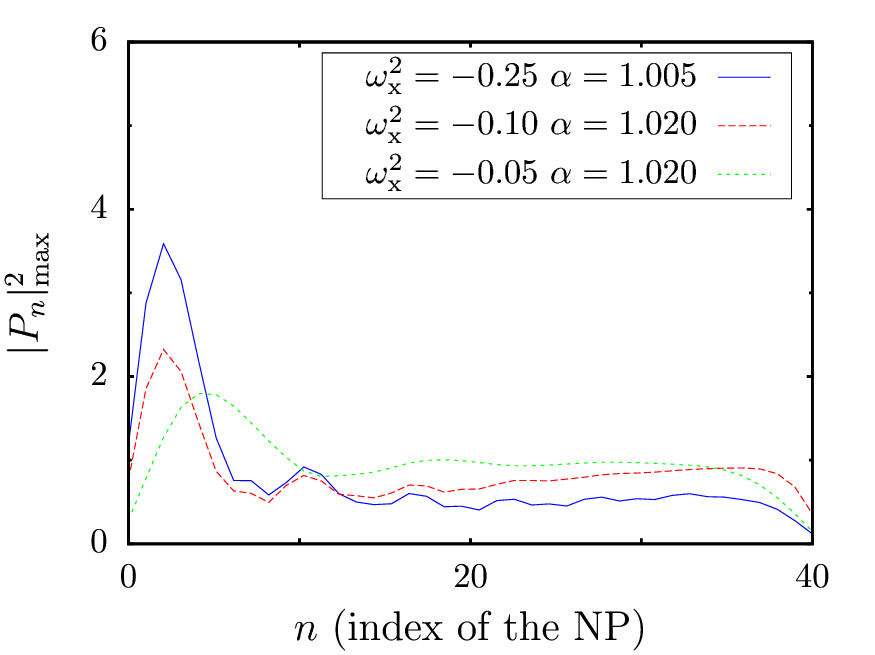}
\caption{Maximum square dipolar moment as function of NP's position for simple graded-chains. The total number of NPs is 41 and $\eta=0.01$.}
\label{FIGsimpleGraded}
\end{figure}

Finally, in this subsection we compared the excitation concentration of mirrored graded-chains with that of simple graded chains. A simple graded-chain corresponds to one half of the system discussed above, equivalent to consider only NPs with indexes from $0$ to $(N-1)/2$ in Fig. \ref{system} for example.
Fig. \ref{FIGsimpleGraded} shows excitation concentration in terms of the maximum value of $|P_i|^2$ vs NP's positions for a sufficiently large chain. The figure can be compared with Fig. \ref{FIGP2-N} for large $N$s as the same three pairs of $\omega_{\mathrm{x}}^2$ and $\alpha$ values where used. As expected, simple graded-chains also behave as resonant cavities. However, they present lower excitation concentrations relative to that of mirrored graded-chains, especially for high $\omega_{\mathrm{x}}^2$ values. Furthermore, the position of excitation's maximum along the chains depends on the interaction strength, which makes them harder to control.

\section{CONCLUSIONS.}

 We have studied a kind of nanostructure not previously reported, up to our knowledge, that we called mirrored graded-chain of NPs. We showed that they are able to concentrate or focus light even when system's sizes are negligible compared with the wavelength of the external excitation. The phenomenon was understood by interpreting the system as an effective cavity where plasmonics excitations are trapped between effective band edges, consequence of the change of the passband with NP's positions.

Dependence of excitation concentration on several system parameters was also assessed. Including, different models for distances changing, the degree of changing of distances with position, as well as NP's couplings, damping, and resonant frequencies. A method to build universal curves for the property of interest was developed to understand all these system's parameters. 
The key idea is based on splitting system's response into geometric components, that in general need to be evaluated  numerically, and other system's parameters which will be used to rescale the property of interest.
The method is quite general and can be used on many other situations as in principle it only requires arrays made of identical particles (nano- or not) well described in the quasistatic limit. Note also the similarity between Eq. \ref{matrix} and the resulting equations for a set of classical coupled harmonic oscillators \cite{calvo} or the dynamic of a single particle in quantum mechanics \cite{BCP-01,calvo}.
Anyway, provided the mentioned conditions are fulfilled, the proposed method can result especially useful for situations where system's complexity limits the possibility of analytical solutions but general trends were anyhow necessary.

There are a number of applications where mirrored graded-chains of NPs can be useful. In the case of sensors, it can help to solve the problem of hotspot's volume vs intensity. 
It is known that electromagnetic fields become more intense inside hotspots when the curvature of a NP is more pronounced or when we stack increasingly small NPs in what is known as nanolenses. However, the volume that enclose these hotspots is reduced at the same time and this effect can be so drastic that not even a single molecule can fit into the hotspot. With our proposal, we are increasing the electromagnetic fields not by sharpening a single NP or using nanolenses, but by tuning appropriately the interactions among NPs. Another possibility is to use our system for multi-frequency simultaneous sensing as it possessed several resonances physically separated.

Mirrored graded-chains can also be used for near field spectroscopies without an actual tip. In this case the system would act as an effective tip where electromagnetic fields are much higher when the analyte is close to the center of the array and negligible if the analyte is moved away from it. Alternatively it can be used to calibrate more standard near-field-spectroscopic devices.

Its application to solar cells would require a more complex design of the array as one generally need to concentrate the electromagnetic fields on a 3D volume. One possibility is to use arrays made of multiple 2-D layers where the interaction between layers changes gradually and the active medium surrounds the NPs that form the central layers \cite{AtwaterPolmanRev}. Alternatively, one can use the direct light-to-heat conversion mechanism mediated by conductive nano-particles.
The dramatic light concentration produced by nanostructures such as the one proposed here, induces vaporization of the host medium without the requirement of heating the liquid volume to the boiling point \cite{NordlanderLight-Heat}.

Finally, one can also use the proposed structure for plasmon-enhanced photo-detectors, helping to reduce the size of the photo-detector. This would result in increased speed, decreased noise, and reduced power consumption \cite{BrongersmaRev}.

\section{ACKNOWLEDGEMENTS.}
The authors acknowledge the financial support from CONICET, SeCyT-UNC, ANPCyT,
and MinCyT-C\'{o}rdoba.
\bibliography{./EPB-JPCM-02.bib}
\end{document}